\documentclass{elsart}
\usepackage{graphicx}
\usepackage{amssymb}

\begin{document}
\begin{frontmatter}

\title{Structural study of Cu$_{2-x}$Se alloys produced by mechanical alloying}
\author{J. C. de Lima},
\author{K. D. Machado\corauthref{cor1}},
\ead{kleber@fisica.ufsc.br}
\corauth[cor1]{Corresponding author.}
\author{T. A. Grandi},
\author{A. A. M. Gasperini},
\author{C. E. Maurmann},
\author{S. M. Souza},
\author{C. E. M. Campos}
\address{Depto de F\'{\i}sica, Universidade Federal de Santa Catarina, Florian\'opolis, 
Santa Catarina, Brazil, Cx. P. 476, 88040-900}
\author{A. F. Pimenta}
\address{Depto de F\'{\i}sica, Universidade Estadual de Ponta Grossa, Ponta Grossa, Paran\'a, Brazil, 
84300-900}

\begin{abstract}

The crystalline structures of superionic high temperature copper selenides Cu$_{2-x}$Se ($0 \le x \le 0.25$) 
produced by Mechanical Alloying were investigated using X-ray diffraction (XRD) technique. The measured 
XRD patterns showed the presence of the peaks corresponding to the crystalline superionic high temperature 
$\alpha$-Cu$_2$Se phase in the as-milled sample, and its structural data were determined by means of a 
Rietveld refinement procedure. After a heat 
treatment in argon at 200$^\circ$C for 90 h, this phase transforms to the superionic high temperature 
$\alpha$-Cu$_{1.8}$Se phase, whose structural data where also determined through the Rietveld 
refinement. In this phase, a very low occupation of the trigonal 32(f) sites ($\sim 3$\%) by Cu ions is 
found. In order to explain the evolution of the phases in the samples, two possible mechanisms 
are suggested: the high mobility of Cu ions in superionic phases and 
the intense diffusive processes in the interfacial component of samples produced by Mechanical Alloying.
\end{abstract}

\end{frontmatter}

\section{Introduction}

Due to their physical and chemical characteristics, selenium-based alloys are very important from  
technological and scientific points of view, mainly those alloys containing germanium, zinc or copper 
because of their very interesting optical properties. Recently, cation-deficient copper selenide 
Cu$_{2-x}$Se, which is a mixed ionic-electronic superionic conductor with a homogeneity range 
of $0 \le x \le 0.25$, has been extensively studied because of its interesting properties and potential 
application in solar cells \cite{Lakshmikumar}, window material \cite{Haram}, optical filter 
\cite{Toyoji}, 
superionic conductor \cite{Chen}, electro-optical devices \cite{Kainthla}, thermoelectric 
converter \cite{Rau} and also as a precursor for preparation of copper-indium-diselenide CuInSe$_2$ 
(CIDS) \cite{Ueno}. The superionic transition from $\beta$ phase (low temperature) to 
$\alpha$ phase (high temperature) occurs at 414 K for Cu$_2$Se and decreases with increasing deviation 
from stoichiometry. At room temperature, the superionic $\alpha$ phase is stable for $0.15 \le x \le 0.25$ 
\cite{Abrikosov}. Cu$_{2-x}$Se films are typically p-type, highly conducting, semitransparent, with band 
gap between 1.1 and 1.4 eV, very suitable for solar energy conversion. A solar cell (Schottky type) 
employing a semi-transparent layer of Cu$_{2-x}$Se as window material on n-type semiconductor is estimated 
to have excellent photovoltaic properties and conversion efficiency around 9 \% \cite{Okimura}. 
Many preparation methods of Cu$_{2-x}$Se alloys have been reported including sonochemical syntesis 
\cite{Xu,Xie}, chemical bath deposition \cite{Garcia,Clement,Pathan}, photochemical route 
\cite{Yan}, $\gamma$-irradiation \cite{Qiao}, solid state reaction \cite{Danilkin}, microwave assisted 
heating \cite{Zhu}, hydrothermal method \cite{Wang}, electrodeposition \cite{Lippkow,Battacharya} 
and Mechanical Alloying (MA) \cite{Ohtani}.

The crystalline structure of Cu$_{2-x}$Se has been studied several times 
\cite{Danilkin,Rahlfs,Borchert,Heyding,Oliveira,Boyce,Yamamoto,Montereuil,Milat} but the precise 
distribution of Cu ions and 
the structure itself are still under discussion and diverging crystallographic data can be found in 
the literature. Rahlfs \cite{Rahlfs} and Borchert \cite{Borchert} proposed a structural model for 
$\alpha$-Cu$_{2-x}$Se consisting of a cage of F$\overline{4}$3m symmetry, built by Se atoms in 4(a) 
sites and Cu ions in 4(c) sites, and a mobile cation subsystem formed by the remaining Cu ions  
distributed over the interstitial sites (tetrahedral 4(d), octahedral 4(b), and trigonal 16(e) 
sites). Heyding and Murray \cite{Heyding} considered an fcc cage formed only by Se atoms and a mobile 
subsystem formed by Cu ions in tetrahedral 8(c) and trigonal 32(f) sites, with an overall symmetry 
Fm3m. Neutron diffraction results \cite{Oliveira}, on the other hand, suggest that Cu ions are 
distributed over the trigonal 32(f) sites leaving the octahedral sites empty. The 
$\beta$-Cu$_{2-x}$Se phase are either described as monoclinic or tetragonal \cite{Montereuil,Milat}. 
The ordering of Cu ions in the $\beta$-phase results in a complicated superstructure very sensitive 
to the composition and preparation technique.

In this paper we report results obtained for Cu$_{2-x}$Se alloys produced by the MA technique \cite{MA}. 
MA has been used for almost two decades to produce many unique materials. 
These include, for instance, nanostructured alloys, amorphous compounds and unstable and metastable 
phases \cite{JoaoFeTi,Weeber,Froes,Yavari,CarlosFeSe,KleNiTi,SeZn}. 
This method has several intrinsic advantages, like low temperature processing, easy 
control of composition, relatively inexpensive equipment, and the possibility of scaling up. 
Although the MA technique is relatively simple, the physical mechanisms involved are not yet fully 
understood. In order to make use of this technique in industrial applications, a better understanding 
of these physical mechanisms is desirable.

\section{Experimental procedures}

A binary mixture of high-purity elemental powders of copper (Vetec 99.5\%, particle size $< 10$ $\mu$m) 
and selenium (Alfa Aesar 99.999\% purity, particle size $< 150$ $\mu$m) with nominal composition 
Cu$_2$Se was sealed together with several steel balls into a cylindrical steel vial under an argon 
atmosphere. The ball-to-powder weight ratio was 5:1. A Spex Mixer/Mill model 8000 was used to perform MA 
at room temperature. The mixture was continuously milled for 72 h. A ventilation system was used to keep 
the vial temperature close to room temperature. For this sample, an x-ray diffraction (XRD) measurement 
was performed. After that, it was heat 
treated at 200$^\circ$C for 90 h in quartz capsule containing argon, 
and a new XRD measurement was collected. 
All XRD measurements were recorded on a Siemens diffractometer with a graphite monochromator in 
the diffracted beam, using the Cu K$_{\alpha}$ line ($\lambda  = 1.5418$ \AA). 

\section{Results and discussion}

\begin{figure}[h]
\begin{center}
\includegraphics{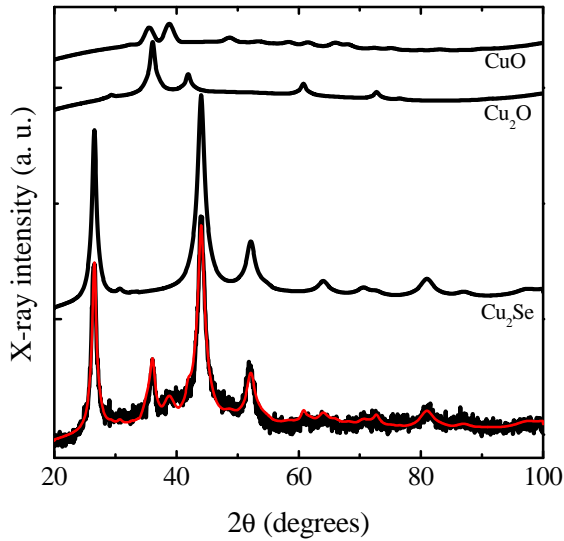}
\caption{\label{fig1} XRD pattern for AM-Cu$_{2-x}$Se and its Rietveld simulation, formed by three phases: 
$\alpha$-Cu$_2$Se, Cu$_2$O and CuO.}
\end{center}
\end{figure}

Figure~\ref{fig1} shows the x-ray diffraction (XRD) pattern taken for the as-milled sample 
(AM-Cu$_{2-x}$Se). A comparison of this pattern with JCPDS cards indicated the 
presence of Cu$_2$Se phase in the F23 structure given by JCPDS card 76-0136. The main peaks of this phase 
are seen at about 2$\theta =26.4^\circ$, $44.0^\circ$, $52.1^\circ$ and $80.9^\circ$, in very good 
agreement with that card. This is a high temperature $\alpha$ phase of the Cu$_{2-x}$Se system, indicating 
that MA performed at room temperature can produce these phases. 
The pattern shown in Fig. \ref{fig1} was simulated using the Rietveld procedure \cite{Rietveld}, and the 
simulation is also seen in this figure. In Ref. \cite{Garcia}, using a chemical bath 
deposition method, Garcia {\em et al.} obtained the Cu$_{1.85}$Se phase corresponding to the mineral 
berzelianite (JCPDS card 06-0680). Danilkin {\em et al.} \cite{Danilkin} produced the low temperature 
monoclinic $\beta$-Cu$_2$Se phase using solid state reactions. 
We also investigated these possibilities but Rietveld refinements using 
these structures did not furnish good results. The lattice parameters obtained for $\alpha$-Cu$_2$Se 
phase was $a=b=c=5.8164$ \AA. The occupation of the sites was also refined, and these data are given 
in Table~\ref{tab1}. 

\begin{table}[h]
\caption{\label{tab1} Rietveld refinement results for $\alpha$-Cu$_2$Se, space group F23, 
JCPDS card 76-0136. The $B_{iso}$ thermal parameter was set to 0 for all sites.}
\begin{tabular}{lccc}\hline
Atom & Site  & Position & Occupation (\%)\\
Cu & 4(c) & $\frac{1}{4},\frac{1}{4},\frac{1}{4}$ & 100\\
Cu & 4(a) & $\frac{1}{2},\frac{1}{2},\frac{1}{2}$ & 42\\
Cu & 16(e) & $\frac{1}{3},\frac{1}{3},\frac{1}{3}$ & 8\\
Cu & 16(e) & $\frac{2}{3},\frac{2}{3},\frac{2}{3}$ & 8\\
Se & 4(a) & $0,0,0$ & 100\\\hline
\end{tabular}
\end{table}

In addition to the peaks of the $\alpha$-Cu$_2$Se phase, the XRD pattern of AM-Cu$_{2-x}$Se has some other 
peaks, which were identified as belonging to the oxides Cu$_2$O (JCPDS card 78-2076) and CuO (JCPDS 
card 80-1917). 
Although the sample was kept under argon atmosphere during the milling, it was probably contaminated 
by oxygen during its preparation and also during the XRD measurement. About 73\% of the crystalline 
phases present in this sample is given by the $\alpha$-Cu$_2$Se phase, whereas the contaminant phases 
are responsible for 18\% (Cu$_2$O) and 9\% (CuO), respectively. The contribution of the three phases 
to the XRD pattern of AM-Cu$_{2-x}$Se is also shown in Fig. \ref{fig1}. 
It should be noted that the strong diffuse scattering seen in Fig. \ref{fig1} is a well-established 
characteristic of the $\alpha$-Cu$_2$Se phase, due to the large number of 
vacancies in the sublattice of Cu atoms \cite{Sakuma}. Thus, we did not consider it as an amorphous part in 
the Rietveld refinement. The refined 
lattice parameters of the oxides are $a=b=c=4.3108$ \AA, for Cu$_2$O (space group Pn$\overline{3}$m), 
and $a=4.6889$ \AA, $b=3.4261$ \AA, $c=5.1322$ \AA\ and $\beta=99.66^\circ$ (space group Cc). The 
average crystallite sizes found using the Scherrer formula \cite{Scherrer} 
for $\alpha$-Cu$_2$Se, Cu$_2$O and CuO are 60 \AA, 78 \AA\ and 50 \AA, respectively, 
showing that all of them are in nanometric form.

\begin{figure}
\begin{center}
\includegraphics{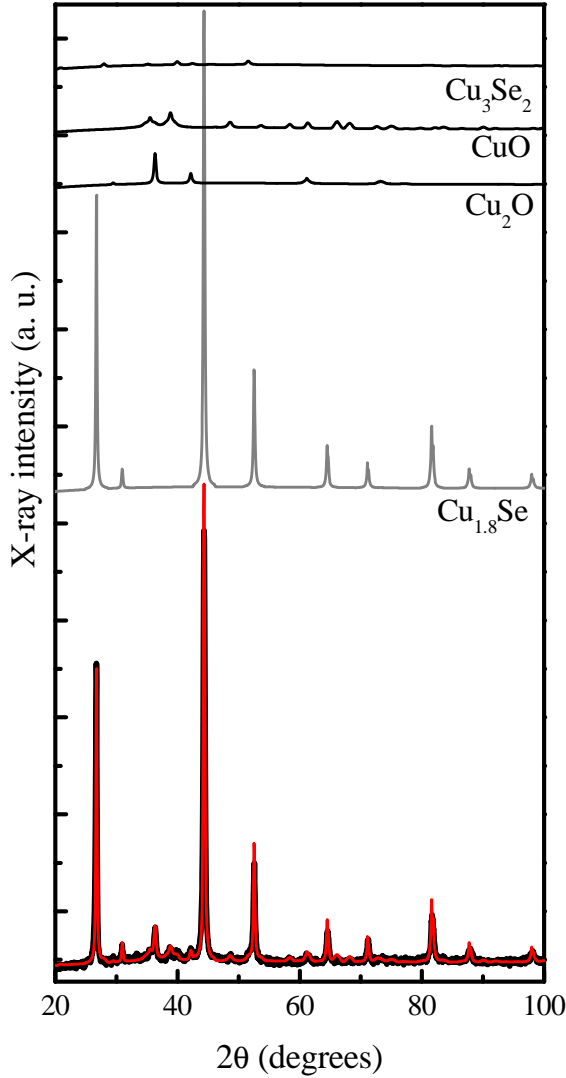}
\caption{\label{fig2} XRD pattern for HT-Cu$_{2-x}$Se and its Rietveld simulation, formed by four phases: 
$\alpha$-Cu$_{1.8}$Se, Cu$_2$O, CuO and Cu$_3$Se$_2$.}
\end{center}
\end{figure}

To study the stability of the $\alpha$-Cu$_2$Se phase, a thermal treatment was made at 200$^\circ$C for 
90 h. The XRD pattern of the heat treated Cu$_{2-x}$Se sample (HT-Cu$_{2-x}$Se) is shown in 
Fig. \ref{fig2}. A comparison with Fig. \ref{fig1} indicates the presence of a phase different from those 
found in AM-Cu$_{2-x}$Se. 
The peaks at about $2\theta = 26.5^\circ$, $44.3^\circ$, $52.3^\circ$, $64^\circ$ and 
$81.7^\circ$ are the most intense of the superionic high temperature phase $\alpha$-Cu$_{1.8}$Se 
given in JCPDS card 71-0044, that corresponds to one of the forms of the mineral berzelianite. This phase 
belongs to the space group Fm$\overline{3}$m, and its refined lattice parameters are $a=b=c=5.7762$ \AA. 
The Rietveld refinement indicated that the Cu ions are found mainly in the tetrahedral 8(c) sites. The 
trigonal 32(f) sites are almost unoccupied. In addition, the refined position of the trigonal 
32(f) sites agrees with those given in the ICSD card 238, as can be seen in table \ref{tab2}, which gives 
the refined data concerning this phase. Danilkin {\em et al.} \cite{Danilkin} have also studied this phase, 
but they found the trigonal 32(f) sites in a different position and with a different occupation. Our 
results agree with the neutron diffraction study of Oliveira {\em et al.} 
\cite{Oliveira}, which did not give any indication of octahedral occupation.

\begin{table}
\caption{\label{tab2} Rietveld refinement results for $\alpha$-Cu$_{1.8}$Se, space group Fm$\overline{3}$m, 
JCPDS card 71-0044.}
\begin{tabular}{lcccc}\hline
Atom & Site  & Position & $B_{iso}$ (\AA$^2$) & Occupation (\%)\\
Cu & 8(c) & $\frac{1}{4},\frac{1}{4},\frac{1}{4}$ & 5.2 & 80\\
Cu & 32(f) & $\frac{1}{3},\frac{1}{3},\frac{1}{3}$ & 1.7 & 3\\
Se & 4(a) & $0,0,0$ & 3.6 & 100\\\hline
\end{tabular}
\end{table}

In addition to the $\alpha$-Cu$_{1.8}$Se phase, the HT-Cu$_{2-x}$Se sample contains the two oxides already 
seen in AM-Cu$_{2-x}$Se, and a small quantity of tetragonal Cu$_3$Se$_2$ phase known as umangite, 
given in JCPDS card 41-1745 (space group P$\overline{4}2_1$m). 
About 75\% of the crystalline phases found in HT-Cu$_{2-x}$Se are given by 
the $\alpha$-Cu$_{1.8}$Se phase, and the contribution of the other phases is 6\% of Cu$_2$O, 16\% of 
CuO and 3\% of Cu$_3$Se$_2$. 
The refined lattice parameters obtained for the contaminant phases are 
$a=b=c=4.2850$ \AA\ for Cu$_2$O, $a=4.6826$ \AA, $b=3.4163$ \AA, $c=5.1570$ \AA\ and 
$\beta= 99.732^\circ$ for CuO and $a=b=6.3835$ \AA\ and $c=4.2596$ \AA\ for Cu$_3$Se$_2$.

With the heat treatment at 200$^\circ$C, the $\alpha$-Cu$_2$Se phase changes to 
the $\alpha$-Cu$_{1.8}$Se phase and the Cu$_2$O phase decreases while CuO increases. These processes 
could be explained by a combination of two factors. First, in the superionic $\alpha$ phase Cu ions 
have a very high mobility, which could be explained by a very small activation energy for jumps of Cu 
between sites. Second, it is well known that MA produces materials with two components, the nanocrystalline 
component, which preserves the crystalline structure of the crystals, and the interfacial region, formed 
by defect centers. The diffusive processes in the interfacial component can be much more intense than 
they are in the nanocrystalline one, and Cu, Se and O atoms belonging to this region could have reacted 
during the heat 
treatment and produced the structural changes seen in the XRD pattern of the heat treated sample. An 
indication that these processes really take place in the alloy is the large increase in the average 
crystallite 
size of the phases. The values obtained are 390 \AA, 225 \AA, 113 \AA\ and 144 \AA, respectively, for 
$\alpha$-Cu$_{1.8}$Se, Cu$_2$O, CuO and Cu$_3$Se$_2$. The crystallinity of all phases 
improves during heat treatment, in particular the crystallinity of the $\alpha$-Cu$_{1.8}$Se phase, 
whose average crystallite size increases more than six times.

\section{Conclusion}

From the results shown above, we conclude that:

\begin{enumerate}
\item The superionic high temperature $\alpha$-Cu$_2$Se phase can be produced by Mechanical Alloying at 
room temperature. It is the majority phase (about 73\%) in the as-milled Cu$_{2-x}$Se sample. During the 
preparation of the sample a contamination by oxygen occurred, and two oxides, Cu$_2$O and CuO, were formed.

\item The $\alpha$-Cu$_2$Se phase present in the as-milled Cu$_{2-x}$Se sample is found in the space 
group F23, and its refined structural parameters, shown in table \ref{tab1}, are a little different 
from those seen in the literature. The oxides are also found in nanometric form. 

\item After a heat treatment in argon at 200$^\circ$C for 90 h, the structure of $\alpha$-Cu$_2$Se 
changes, and the superionic high temperature $\alpha$-Cu$_{1.8}$Se phase is formed. It is found 
in the space group Fm$\overline{3}$m, and its refined structural parameters are found in table 
\ref{tab2}. Almost all Cu ions 
are found in tetrahedral 8(c) sites, and the trigonal 32(f) sites have a very small occupation 
($\sim 3$\%). There is no indication of octahedral occupation, in agreement with results from 
Refs. \cite{Oliveira,Boyce}. The crystallinity 
of all phases are much improved during the heat treatment, in particular the crystallinity of the 
Cu$_{1.8}$Se phase, whose average crystallite size increases by more than six times, from 60 \AA\ to 
390 \AA. There are at least two possible 
reasons for the structural changes occurred in the sample with the heat treatment. 
First, Cu ions in the superionic 
phases have a high mobility since their activation energy for jumps is expected to be very small, 
and the heat 
treatment could furnish this energy. Second, atoms in the interfacial component of the sample produced by 
Mechanical Alloying have also a very high mobility, and the diffusive processes in this component can be 
very intense. The combination of the two processes could explain the structural changes and 
crystallinity improvement occurred during the heat treatment, and also the formation of the superionic 
phase at room temperature.

\end{enumerate}

\ack

We thank to the Brazilian agencies CNPq and CAPES for financial support.


\end{document}